\documentclass[12pt,tightenlines,preprintnumbers]{revtex4}
\RequirePackage{amsfonts}
\RequirePackage{amssymb}
\RequirePackage{amsmath}
\RequirePackage{graphicx}
\RequirePackage{makeidx}
\RequirePackage{ifthen}
\RequirePackage{color}
\renewcommand{\a}{\ensuremath{\alpha}}
  
\newcommand{\g}{\ensuremath{\gamma}  }

\newcommand{\ve}{\ensuremath{\varepsilon}}

\renewcommand{\th}{\ensuremath{\theta} }

\newcommand{\m}{\ensuremath{\mu}     } 
\newcommand{\n}{\ensuremath{\nu}     }
\newcommand{\s}{\ensuremath{\sigma} }

\newcommand{\G}{\ensuremath{\Gamma}  }

\newcommand{\be}{\begin{eqnarray}}          \newcommand{\ee}{\end{eqnarray}}
\newcommand{\ba}{\begin{align}}          \newcommand{\ea}{\end{align}}
\newcommand{\benum}{\begin{enumerate}}   
\newcommand{\eenum}{\end{enumerate}}
\newcommand{\bitem}{\begin{itemize}}          
\newcommand{\eitem}{\end{itemize}}

\newcommand{\nn}{\nonumber \\ }
\newcommand{\abs}[1]{\left| #1 \right|}   
\newcommand{\abssq}[1]{\left| #1 \right|^2} 
\newcommand{\sci}[2]{#1\times 10^{#2}}  
\newcommand{\psci}[2]{(#1)\times 10^{#2}}  
\newcommand{\fr}[2]{\ensuremath{\frac{#1}{#2}}} 
\newcommand{\pfr}[2]{\left( \frac{#1}{#2}\right)}

\newcommand{\sqrtfr}[2]{\sqrt{\frac{#1}{#2}}} 
\newcommand{\invsqrt}[1]{\frac{1}{\sqrt{ #1}}}     

\newcommand{\bra}[1]{\ensuremath{ \left\langle #1 \right| } }  
\newcommand{\ket}[1]{\ensuremath{ \left| #1 \right\rangle} }  
\def\gfive{\gamma_{5}}
\newcommand{\ctr}[3]{#1_{#3} #2^{#3}}

\newcommand{\usemarker}{Y}
\newcommand{\marker}[1]{
       \ifthenelse{\equal{\usemarker}{Y}}
                     {\vskip 20pt\mbox{}\marginpar{\bf\sf #1}}{}
               }
\newcommand{\mk}[1]{
\ifthenelse{\equal{\usemarker}{Y}}
{\noindent\hskip -1truecm {\bf\blue$^{#1}$}}{}
}
\newcommand{\myname}{Yu-Feng Zhou}
\newcommand{\myaddress}{Ludwig-Maximilians-University Munich, \\Sektion Physik. Theresienstra$\beta$e 37, D-80333. Munich, Germany}
 
\begin{document}
\title{The gluonic $B$  and $J/\psi$ decays into the $\eta'$ meson}
\author{Harald Fritzsch and \myname}
\affiliation{\myaddress}
\date{\today}
\begin{abstract}
The inclusive and exclusive $B$ decays into the $\eta'$ meson plus others are investigated
in a model based on the QCD anomaly. The invariant mass distribution is discussed.
The constraint of the effective coupling is obtained from the data of the exclusive decay 
modes. The branching ratio of $J/\psi\to \eta'\eta \g$ is predicted to be as large as 
$5.4\times 10^{-5}$,
which can be tested in the forthcoming CLEO-c experiments.
\end{abstract}
\preprint{LMU 20-03}
\preprint{hep-ph/0301038} 
\maketitle

{\noindent\bf 1. Introduction.\\} 
A few years ago, unexpected large branching ratios of $B$ decaying
into final states with an $\eta'$ meson such as $B\to \eta' X_{s}$ and $B\to \eta' K$ were observed
by the CLEO collaboration\cite{Browder:1998yb,Behrens:1998dn} and recently confirmed by
BaBar\cite{Aubert:2001sr} and Belle\cite{Belldata}.  This stimulated many theoretical activities in
understanding the special role of the $\eta'$ meson in B decays.
As the contribution of traditional four quark operators from the effective Hamiltonian in the
standard model (SM) is far below the data\cite{Datta:1998nr,Ali:1998nh}, various exotic mechanisms
were introduced such as a large coupling between the gluon and $\eta'$ through the QCD
anomaly\cite{Atwood:1997bn,Hou:1998wy,Fritzsch:1997ps,Gronau:1996ng,Dighe:1997hm}, intrinsic $\bar c c$ content inside
$\eta'$\cite{Halperin:1998ma,Yuan:1997ts} and positive interference between several contributions in
the SM\cite{Lipkin:1991us,Beneke:2002jn} et.al. The large contribution arising from new physics was
also discussed\cite{kagan:1997qn,Hou:1998wy}.
Among those theoretical efforts, the possible enhancement from the QCD anomaly is of particular
interest since it is well known that the $\eta'$ meson plays an very special role in the dynamics of
low energy QCD \cite{Fritzsch:1975tx}.

The mass eigenstates $\eta'$ and $\eta$ are related to flavor
octet and singlet states $\eta_{8}$, $\eta_{0}$ through a mixing matrix: 
%
\begin{align}
\eta=\eta_{_8}\cos\th   -\eta_{_0}\sin\th  ,\qquad
\eta'=\eta_{_8}\cos\th  +\eta_{_0}\sin\th ,
\end{align}
%
where $\th$ is the mixing angle and $\eta_{8}$, $\eta_{0}$ have the
flavor content:
$\eta_{8}=\invsqrt6(\bar u u+\bar d d- 2\bar s s)$ and
$\eta_{0}=\invsqrt3(\bar u u+\bar d d+\bar s s)$.
The associated axial currents are 
$j^{\m8}_{5}=\invsqrt6(\bar u\g^{\m}\gfive u+\bar d\g^{\m}\gfive d-2\bar s\g^{\m}\gfive s)$
and 
$j^{\m0}_{5}=\invsqrt3(\bar u\g^{\m}\gfive u+\bar d\g^{\m}\gfive d+\bar s\g^{\m}\gfive s)$
respectively. 
Through the  QCD anomaly, the divergence of the flavor singlet axial current is non-zero and is given by
\begin{align}
   \partial_{\m} j^{\m0}_{5}= 
  \fr{1 }{\sqrt3}\left(2i \sum_{q=u,d,s} m_{q}\bar q \gfive q 
    +\fr{3\a_{s}}{4\pi}\ctr{G}{\tilde G}{\m\n}
    \right)
\end{align}
%
where $\tilde G^{\m\n}=\frac12 \ve^{\m\n\rho\s}G_{\rho\s}$ is the dual of $G_{\m\n}$.
This breaks the global $U(1)_{A}$ symmetry for massless particles and makes the
flavor singlet state $\eta_{0}$ evade to be a Goldstone boson of chiral
$SU(3)_{L}\otimes SU(3)_{R}$ symmetry.  The QCD anomaly gives its main contribution
to the large mass of $\eta' \ (m_{\eta'}=0.958\mbox{GeV})$ which is much heavier
than the other  flavor octet states such as $\pi$, $K$s and suggests a large gluon content in 
$\eta'$.

The QCD anomaly indicates a strong coupling between $\eta'$ and the gluon field. It is then natural
to understand the large branching ratio of $B\to \eta' X_{s}$ through the QCD anomaly. In the
literatures there are two different ways to handle this problem.
The one is through two-body decay process $b\to s \eta'$ from some effective Hamiltonian due to QCD
anomaly\cite{Fritzsch:1997ps}.
The other one is through three-body process $b\to s g \eta'$\cite{Atwood:1997bn,Hou:1998wy}.  In the
first step decay, the $b$ quark decays into the $s$ quark and a virtual gluon $g^{*}$, then $g^{*}$
decays into $\eta'$ and a on shell gluon $g$. 
This model   has some advantages
in explaining  the spectrum of invariant mass distribution of recoiling hadrons. However, 
the effective $gg^{*} \eta'$ vertex  seems  to be too small from various approaches
\cite{Hou:1998wy,Muta:1999tc,Ali:2000ci,Artuso:2002px}. 
In both of the approaches, the effective coupling between $\eta'$
and gluon may contain complicate non-perturpative quark-gluon interactions.  It
is then better to treat  it as a free  phenomenological parameter rather than to
evaluate it from perturbative calculations\cite{Fritzsch:1997ps}.

In this note we focus on the phenomenological analysis of the first mechanism.  
The effective Lagrangian  in this model is given by \cite{Fritzsch:1997ps}
\begin{align}\label{Hamiltonian}
H_{eff}=a \a_{s}G_{F} \bar s_{L}b_{R} \ctr{G}{\tilde G}{\m\n}+ \mbox{h.c}
\end{align}
%
where $\a_{s}$ and $G_{F}$ are the strong coupling constant and Fermi constant,
$a$ is the effective coupling. 
 From this effective Hamiltonian, the decay $B\to
\eta' X_{s}$ arises from the subprocess $b\to s\eta'$. The evaluation of matrix
elements is straightforward:
%
\begin{align}
\bra{s \eta'}H_{eff}\ket b
=a \a_{s}G_{F}\bra s \bar s_{L}b_{R}\ket b \bra{\eta'}\ctr{G}{\tilde G}{\m\n}\ket0
\end{align}
%
Applying  relation
$\bra0 j^{\m8(0)}_{5}\ket{\eta_{8(0)}}=i f_{8(0)}P^{\m}$  
  to the divergences of both flavor singlet and octet axial
currents and ignoring small $u,d$ quarks masses, the matrix elements
$\bra{0}\ctr{G}{\tilde G}{\m\n}\ket{\eta'}$ can be rewritten as   
\begin{align}\label{gluonContent}
\bra{0}\a_{s}\ctr{G}{\tilde G}{\m\n}\ket{\eta'}
&=\fr{4\pi}{3}\sqrtfr32 m_{\eta'}^{2} (f_{8}\sin\th +\sqrt2 f_{0}\cos\th)
\nn
\bra{0}\a_{s}\ctr{G}{\tilde G}{\m\n}\ket{\eta}
&=\fr{4\pi}{3}\sqrtfr32 m_{\eta}^{2} (f_{8}\cos\th-\sqrt2 f_{0}\sin\th)
\end{align}
%
%
%
In the $b$ quark rest frame, the decay branching ratio is given by
\begin{align}\label{inclusive}
Br(B\to \eta' X_{s})=\fr{\pi}{12} \tau_{B}
a^{2}  G_{F}^2 m_{\eta'}^4  (f_{8}\sin\th +\sqrt2 f_{0}\cos\th)^2
\fr{ (m_{b}^2-m_{\eta'}^2)^2}{m_{b}^3}
\end{align}
\begin{align}\label{inclusive2}
Br(B\to \eta X_{s})=\fr{\pi}{12} \tau_{B}
a^{2}  G_{F}^2 m_{\eta}^4  (f_{8}\cos\th -\sqrt2 f_{0}\sin\th)^2
\fr{ (m_{b}^2-m_{\eta}^2)^2}{m_{b}^3}
\end{align}
where $\tau_{B}$ is the lifetime of $B$ meson. 

\vskip 1cm {\noindent\bf 2. Recoil mass distribution\\} 

For two-body like
subprocess such as $b\to \eta' s$, the invariant mass is directly related to the
energy $E_{\eta'}$ of the $\eta'$ meson through the relation: 
$  m_{X}^{2}=m_{B}^2+m_{\eta'}^2-2 m_{B} E_{\eta'}$,
where $m_{B}$ and $m_{\eta'}$ are the masses of $B$ and $\eta'$ meson.  The
small $s$ quark mass has been ignored.  In the two-body decay of $b\to \eta' s$, the
energy of $\eta'$ is fixed from energy-momentum conservation. A typical value of
the pole mass $m_b=4.8$ GeV will lead to a narrow peak with the invariant mass of
$m_{X_{s}}\simeq 1.5$ GeV.
This seems to be disfavored by the current data since the experiment reported a
peak at about 2 GeV with a relative wide width in the recoil mass distribution
\cite{Browder:1998yb,Aubert:2001sr}.

However, the above estimation may be too naive. Note that in the two-body decay
process, the exact distribution of the recoil mass strongly depends on the wave
function of the $B$ meson which is theoretically hard to estimate. It is too early
to draw the conclusion that the current data already disfavored all the two-body
models.

\begin{figure}[hbt]
\begin{center}
\includegraphics[width=0.4\textwidth]{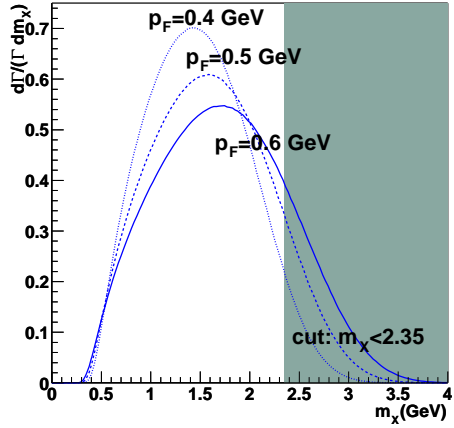}
\caption{Recoil mass distribution from process $b\to s \eta'$ with Fermi motion being included.
  The solid, dashed and dotted curves correspond to $p_{_{F}}=$0.6, 0.5 and 0.4 GeV. The
value of $m_{q}$ is fixed at $0.15$ GeV.
  The shadowed area indicates the acceptance cut of $m_{X}<2.35$ GeV from the
  CLEO experiment}
\label{FermiMotion} 
\end{center}
\end{figure}

To illustrate  the non-perturbative bound state effects 
here we adopt a simple model proposed by Altarelli
et.al.  \cite{Altarelli:1982kh} a number of years ago which 
is based on the Fermi motion of the $b$ quark inside $B$ meson.
The basic idea of this model is that the Fermi motion of the $b$ quark and the spectator
quark $q$ in the B meson make them have back-to-back relative three-momenta
$\mathbf{p}$ in the $B$ rest frame. The momentum is assumed to obey a Gaussian
distribution as follows
\begin{align}
\phi(p)=\fr{4}{\sqrt{\pi} p_{_F}^{3}} e^{-p^2/p_{_F}^2}, \quad p=\abs{\mathbf{p}}
\end{align}
%
where $\phi(p)$ is normalized as
$\int_{0}^{\infty} \phi(p) p^{2} dp=1$.
The mean value of $p$ is $<p>=\fr32 p_{_F}$.  In this model the spectator quark $q$
is on always handled as on shell while the $b$ quark is treated as off-shell.
Through energy-momentum conservation, the effective mass $W$ of the b quark is
determined as
\begin{align}
W^{2}&=m_{B}^2+m_{q}^2-2 m_{B}\sqrt{m_{q}^2+p^2}
\end{align}
%
and the energy of the $b$ quark is $E_{W}=\sqrt{W^2+p^2}$. Here one parameter $p_{_F}$ is
introduced which specifies both the distribution width and the mean value.  As
$p_{_F}$ is linked to the average energy of $b$ quark inside the  $B$ meson, in
principle it can be calculated from theories based on non-perturbative methods or
from some models. For example, the calculations from QCD sum rule
\cite{Bagan:1995qw} give $p_{_{F}}=0.58\pm0.06$ GeV, and the value from relativistic
quark model \cite{Hwang:1995vz} is $0.54\pm0.16$ GeV. The value of $p_{_F}$ can also
be extracted directly from the data.  A fit to $B\to X_{s}\g$ photon energy
spectrum give a value of about $0.45$ GeV\cite{Ali:1995bi} while the fits to semi
leptonic $B$ decays and $B\to J/\psi X$ give a value of 0.57 GeV
\cite{Palmer:1997wv,Hwang:1996ie}. Thus the value of $p_{_{F}}$
is likely to lie in the range $0.4 \lesssim p_{_{F}} \lesssim 0.6$ GeV.
After including the Fermi motion, the differential decay width $d \G(m_{b})/ dm_{X}$, should
be replaced by
\begin{align}\label{mxdistr}
\fr{d\G}{dm_{X}}= 
\int_{0}^{p_{max}} dp \ \phi(p) p^{2}\cdot 
\fr{d\G(W)}{dm_X}  
\end{align}
where $p_{max}$ is the allowed maximum value of $p$, $d\G(W)/dm_{X}$ is the
differential decay rate in the $B$ meson rest frame, which is linked to the one in
$b$ quark rest frame through a Lorentz boost\cite{Palmer:1997wv,Du:1998ft}
 In Fig.\ref{FermiMotion} the invariant mass distribution is generated in this model
with different values of $p_{_{F}}$. Here we 
use  $\fr{1}{\G}\fr{d\G}{dm_{X}}$ which is normalized to unity and independent on 
the value of $a$.
It can be clearly seen that the $p_{_F}$ dependence is rather strong.  The peak of the
distributions significantly shifts from around $\simeq 1.4$ GeV ( for
$p_{_F}=0.4$ GeV) to $\simeq 1.8$ GeV (for $p_{_F}=0.6$ GeV ). Considering the
considerable uncertainties in both theory and experiment data, there is no
significant disagreement in the recoil mass distribution of $B\to \eta' X_{s}$.
 
\vskip 1cm
{\noindent\bf 3. Bound on $a$ from inclusive and exclusive $B$ decays\\}
The value of $a$ could  be constrained from  the exclusive decay modes
$B\to \eta'(\eta) K^{(*)}$. 
Note that although predictions of the standard effective Hamiltonian approach  
are too low to account for the data of inclusive decay modes, the disagreement
 in the exclusive decay modes are smaller\cite{Ali:1998nh,Beneke:2002jn}. 
Furthermore, the effective Hamiltonian approach can reproduce the correct  patterns of 
$Br(B\to\eta'K) \gg Br(B\to\eta K)$  and $Br(B\to \eta'K^{*})\lesssim Br(B\to \eta K^{*})$ which is observed in the experiments.
It implies that in exclusive decays modes, it may still  play an important role, and
the interference between different contributions may also be important\cite{Chiang:2001ir,Gronau:1999hq}.

Nevertheless, by saturating the current data of exclusive decays, the upper bound of the 
parameter $a$ can still
be obtained.
 The decay amplitudes of decay modes $B\to \eta' K^{(*)}$
and $B\to \eta K^{(*)}$ in this model read  
\begin{align}
\mathcal{M}(B^{\pm,0}\to\eta' K^{\pm,0})&=a  G_{F} \fr{4\pi}{3 }\sqrtfr32 m_{\eta'}^{2}(f_{8} \sin\th +\sqrt2 f_{0}\cos\th)
      \fr{m_{B}^2-m_{K}^2}{2(m_{b}-m_{s})}
F_{0}^{BK}(m_{\eta'}^2)
\nn\nn
\mathcal{M}(B^{\pm(0)}\to\eta K^{\pm(0)})&=a  G_{F} \fr{4\pi}{3 }\sqrtfr32 m_{\eta}^{2}(f_{8} \cos\th -\sqrt2 f_{0}\sin\th)
      \fr{m_{B}^2-m_{K}^2}{2(m_{b}-m_{s})}
F_{0}^{BK}(m_{\eta}^2)
\nn
\nn
\mathcal{M}(B^{\pm(0)}\to\eta' K^{*\pm(0)})&=-a G_{F} \fr{4\pi}{3}\sqrtfr32 
m_{\eta'}^{2}(f_{8} \sin\th +\sqrt2 f_{0}\cos\th)
\fr{ \abs{P_{\eta' K^*}}m_{B}}{(m_{b}+m_{s})}    
A_{0}(m_{\eta'}^2)
\nn
\nn
\mathcal{M}(B^{\pm(0)}\to\eta K^{*\pm(0)})&=-a G_{F} \fr{4\pi}{3}\sqrtfr32 
m_{\eta}^{2}(f_{8} \cos\th -\sqrt2 f_{0}\sin\th)
\fr{ \abs{P_{\eta K^*}}m_{B}}{(m_{b}+m_{s})}    
A_{0}(m_{\eta}^2)
\end{align}
where $\abs{P_{\eta' K^*}}\simeq \abs{P_{\eta K^*}}\simeq \fr12 m_{B}$. 
$F^{BK}_0(q^2)$ and $A_0(q^2)$ are the form factors for $B\to K$ and $B\to K^*$
transition with momentum transfer $q^2$.
The value of $m_{b}$ is taken to be the effective one. i.e 
$m_{b}^{-2}\simeq\int W^{-2} \phi(p)p^{2} dp$. In the calculations, we take $m_{b}=4.65$ GeV 
which corresponds to  $p_{_{F}}=0.5$GeV and  $m_{q}=0.15$GeV.

The corresponding branching ratio can be evaluated through the following relation
\begin{align}
Br=\fr{\tau_{B} \abs{P}}{8\pi m_{B}^2} \abs{\mathcal{M}}^2
\end{align}
where $\abs{P}$ is the momentum of one of the final state mesons in $B$ rest frame.


It is useful to define two kind of  ratios which are independent of the parameter $a$:

1) The ratio between $B\to\eta'X$ and $B\to\eta X (X=X_{s}, K or K^{*})$. This ratio
is independent of the value of $a$ and only sensitive to the $\eta'-\eta$
mixing. In this model one finds\cite{He:1998xk}
\begin{align}\label{R}     
R&\equiv\fr{Br(B\to \eta X_{s})}{Br(B\to\eta' X_{s})}
=
\fr{Br(B\to \eta  K)}{Br(B\to\eta'  K)}
=
\fr{Br(B\to \eta  K^{*})}{Br(B\to\eta'  K^{*})}
\nn
&=
\fr{m_{\eta}^4}{m_{\eta'}^4}
\left(  \fr{f_{_8}\cos\th-\sqrt{2}f_{_0}\sin\th }{f_{_8}\sin\th+\sqrt{2}f_{_0}\cos\th }\right)^2
\end{align}
In the following numerical calculations we take   
$\th=-17^{\circ}$ and $f_{8}=f_{0}=1.06 f_{\pi}$ \cite{Ball:1996zv} as an illustration. 
This leads to a value of $R\simeq 0.16$. Considering the CLEO data of $R\lesssim (0.1\sim 0.8)$
\cite{Behrens:1998dn}, 
it follows that with the constraints from $B\to \eta K$, this model can account
for at most $\sim 60\%$ of the observed $B\to \eta' K$ branching ratio.  Note
that the exact value of $R$ may vary with different sets of parameters $\th$,
$f_{8}$ and $f_{0}$; the constraints from $R$ are only an order of magnitude estimate.

\begin{figure}[htb] 
\begin{center}
\includegraphics[width=\textwidth]{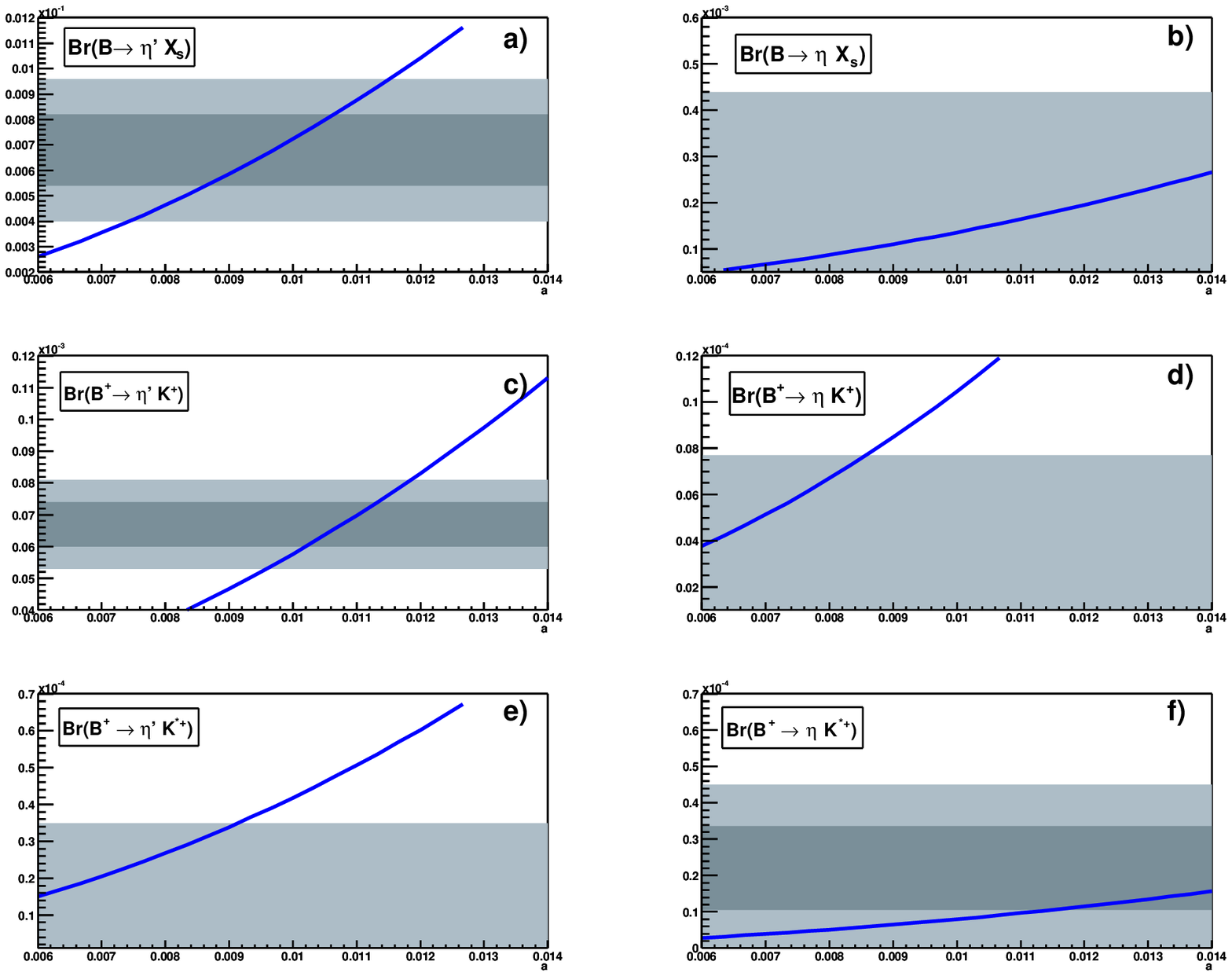}
\caption{ 
Branching ratios for inclusive and exclusive decay modes as a function of $a$.\\
a) For decay mode $B\to \eta' X_{s}$, the dark and
light shadows represent the $1\s(2\s)$ allowed ranges by current data.\\
b) For decay mode $B\to \eta X_{s}$, the light shadows  represent the $90\%$ allowed range.\\
c) For decay mode $B\to \eta' K^{+}$, the dark and 
light shadows represent the $1\s(2\s)$ allowed ranges.\\
d) For decay mode $B\to \eta K^{+}$, the light shadows  represent the $90\%$ allowed range.\\
e) For decay mode $B\to \eta' K^{*+}$, the light shadows  represent the $90\%$ allowed range.\\
f) For decay mode $B\to \eta K^{*+}$, the dark and
light shadows represent the $1\s(2\s)$ allowed ranges.
}
\label{model}
\end{center}
\end{figure}

2) The  ratio between $B\to P K^{*}$ and $B\to P K \ (P=\eta'\ or\  \eta)$. 
In this model it is independent of $both$ the value of $a$ and details of $\eta'-\eta$ mixing.
\begin{align}
R'\equiv\fr{Br(B\to \eta'  K^{*})}{Br(B\to\eta' K)}
=
\fr{(m_{B}^2+m_{K^{*}}^2-m_{\eta}^2)^2-4 m_{B}^2 m_{K^{*}}^2}{(m_{B}^2-m_{K}^2)^2}
\cdot \pfr{A_{0}(m_{\eta'}^2)}{F_{0}^{BK}(m_{\eta'}^2)}^{2}
\end{align}
The values of $F_{0}$ and $A_{0}$ in the  BSW model \cite{Bauer:1987bm} are  $F_0=0.38, A_0=0.32$ which
corresponds to $R=0.84$, while from light cone QCD sum rule
\cite{Ali:1994vd,Ball:1998tj} $F_0=0.35\pm0.05,A_0=0.39\pm0.1$ and $R=1.1\pm
0.3$.  Thus if this model gives the dominant contribution to these modes, the
value of $R$ should be around 1. However, the current data gives a value of $R'
\lesssim (0.5\sim 0.4)$ \cite{Behrens:1998dn}. This is a more clear  and  stronger constraint than the 
one from $R$.
With the observed small value of $R'$, this model can explain at most  half of
the branching ratio of $B\to\eta'(\eta) K$ and therefore is not the dominant mechanism
of these processes.
%
In Fig.\ref{model}(c-f)
the numerical results of branching ratios as a function of the effective coupling $a$
is given and compared with the  data. As some inclusive decay modes 
have not yet been observed by   the  Babar and Belle collaborations,  
 only the CLEO data are used in the numerical evaluations.
It can be seen from the figure that
the data of exclusive decay modes  $B\to \eta' K^{*}$ and $B\to \eta K$ impose strong
constraints on the effective coupling.
With these constraints, the maximum value of $a$  lies in the range:
\begin{align}\label{a-bound}    
a \lesssim \psci{8\sim9}{-3} \quad \mbox{GeV$^{-1}$}
\end{align} 

From  Eq.(\ref{inclusive}) and (\ref{inclusive2}), the branching
ratio of inclusive decays $B\to \eta'(\eta) X_{s}$ as a function of $a$ is plotted
in Fig.\ref{model}(a-b)   and compared with the data. In the decay $B\to \eta' X_{s}$ the acceptance
cut effects is taken into account, which leads to a 19$\%$ reduction from the
calculation in Eq.(\ref{inclusive}). Given  the upper bound of $a$ in Eq.(\ref{a-bound}) 
 this model can still successfully reproduce the $B\to \eta' X_{s}$ branching ratio 
within 1$\s$ range.

\vskip 1cm
{\noindent\bf 4. Prediction of radiative decay $J/\psi\to \eta'\eta \g$\\}
From the effective Hamiltonian in Eq.(\ref{Hamiltonian}), this model can 
also contribute to the radiative  $J/\psi$ decays into $\eta'$. Using relation
 Eq.(\ref{gluonContent}), the ratio between $J/\psi\to \eta' \g$ and 
$J/\psi\to \eta \g$ can be predicted and found to be the same as in Ref.\cite{Ball:1996zv} 
\begin{align}
\fr{\G(J/\psi \to \eta' \g)}{\G(J/\psi \to \eta \g)}
=\abssq{\fr{\bra0 G\tilde G \ket{\eta'}}{\bra0 G\tilde G \ket\eta}}\cdot
\fr{(1-m^{2}_{\eta'})^3}{(1-m^{2}_{\eta})^3}
\end{align}
which is in good agreement with the data.

Furthermore, given the value of the effective coupling $a$ 
the decay rate of  $J/\psi\to \eta'\eta \g$ can be predicted.
To this end let us first define the ratio
\be\label{ratio}
r(\eta')&=&\G(B\to \eta' X_{s})/\G(B\to g^{*}X_{s})
\ee
which can be understood as the size of $b\to s\eta'$ relative to  $b\to s g$. 
Taking   $\G(B\to g^{*} X_{s}) \sim 1\%$ and  $a \lesssim 0.008$ GeV$^{-1}$ 
which comes from the bounds from exclusive decays
as an example, one finds  
\begin{align}  
r(\eta')\lesssim 0.045
\end{align}
Note that the strong coupling constant in the effective Hamiltonian has been
separated from the effective coupling $a$ and 
absorbed in the matrix element of $\bra0 \a_{s}G\tilde{G}\ket{\eta'(\eta)}$. It is
expected that there is no significant running on the value of $a$ from energy
scale $m_{B}$ to $m_{J/\psi}$.
 
Since the radiative decay of $J/\psi \to \g X$ is dominated by the process 
$J/\psi\to  g^{*}g^{*} \g$,    the branching ratio of $J/\psi\to \eta'\eta' \g$
can be simply estimated as 
\be
\fr{Br(J/\psi\to \eta'\eta' \g)}{Br(J/\psi \to \g X)} \simeq  r(\eta')^2.
\ee
The observation of the  process $J/\psi \to \g X$ gives
$Br(J/\psi \to \g X)=(17.0\pm 2.0)\times 10^{-2}$. Thus 
taking $r(\eta')=0.045$ the maximum   branching ratio
of   $J/\psi\to \eta'\eta' \g$ is estimated as 
\be
Br(J/\psi\to \eta'\eta' \g) \simeq  \sci{3.4}{-4} 
\ee
The decay rate of $J/\psi \to \eta'\eta \g$ can be estimated as follows
\begin{align}
\fr{Br(J/\psi \to \eta'\eta \g)}{Br(J/\psi \to \eta'\eta' \g)}
=R
\end{align}
Using the value of $R=0.16$ from Eq.(\ref{R}) one finds for 
the maximum branching ratio for $J/\psi \to \eta'\eta \g$  
\begin{align}
Br(J/\psi \to \eta'\eta \g) \simeq \sci{5.4}{-5}
\end{align}

Considering the detection efficiency of $\eta'$ is about a few percent
( through $\eta'\to \eta \g \g$), it may be hard to find a signal of such 
a decay mode in BES due to limited statistics ( in BES $\sci{5}{7} J/\psi$ samples
are collected). But in the  forthcoming  CLEO-c project $\sci{1}{9} J/\psi$ samples
are planned to be produced. It is then promising to search for the signal and test the
predictions from this model in the CLEO-c experiment.

\begin{acknowledgments}
Y.F. Zhou acknowledges the support by the  Alexander von Humboldt Foundation.
\end{acknowledgments}
    
\bibliographystyle{apsrev}
\bibliography{reflist}

\end{document}